\documentclass[preprint,showpacs,preprintnumbers,amsmath,amssymb]{revtex4}
\usepackage{graphicx}
\usepackage{dcolumn}
\usepackage{bm}
\begin{document}
\title{Dielectric function with exact exchange 
contribution in the electron liquid. II. Analytical expression}
\author {Zhixin Qian}
\affiliation{Department of Physics,
Peking University, Beijing 100871, China}
\date{\today}

\begin{abstract}

The first-order, in terms of electron-interaction in the
perturbation theory, of the proper linear response function 
$\Pi ({\bf k}, \omega )$ gives rise to the exchange-contribution to the
dielectric function $\epsilon ({\bf k} , \omega)$ in the electron liquid. 
Its imaginary part, $Im \Pi_1 ({\bf k}, \omega)$, is calculated exactly. 
An analytical expression for $Im \Pi_1 ({\bf k}, \omega)$
is derived which after refinement has a quite simple form.

\end{abstract}

\maketitle

\section{Introduction with concluding remarks}

Electronic excitations are one of major 
subjects in solid state physics \cite{Pines};
the dielectric function $\epsilon ({\bf k} , \omega)$ of
the homogeneous electron liquid \cite{Fetter,Mahan,Pines1} has been playing
a central role in the description of these excitations. 
In the preceding paper \cite{Qian}, referred to as I hereafter, the
static dielectric function $\epsilon ({\bf k} , 0)$ with exchange
contribution was studied. 
A very simple
expression for $\Pi_1 ({\bf k}, 0)$ the first order, in terms of  
electron-interaction in the perturbation
theory, of the static
proper linear response function $\Pi ({\bf k}, 0)$
in the electron liquid, was derived.
In this paper we set as our task to make 
like development for $\Pi_1 ({\bf k}, \omega)$, its 
dynamical counterpart. An analytical expression is obtained
for $Im \Pi_1 ({\bf k}, \omega)$,  the imaginary part 
of $\Pi_1 ({\bf k}, \omega)$.

The conceptual importance of
$\epsilon ({\bf k} , \omega)$ [and $\Pi ({\bf k}, \omega)$]
and previous progress made in the study of them have been briefly 
introduced in I, with emphasis on their static aspect.
In general previous works in both of experimental and theoretical 
respects are enormous.
We here limit ourselves to mentioning several of them which bear most close 
theoretical relation to the present paper 
\cite{Bohm,Lindhard,Hubbard,DuBois,Nozieres,Osaka,Glick,Glick2,
Ninham,DuBois1,Kleinman,Langreth,
Hasegawa,Toigo,Rasolt,Rajagopal,Niklasson,Holas1,Holas2,
Brosens,Tripathy,Dharma,Awa,Holas3,Gasser,Richardson,Nifosi,Vignale}. 
Particularly noteworthy is the work by Holas et al in Ref. \cite{Holas1}
in which an analytical 
expression for $Im \Pi_1 ({\bf k}, \omega)$ 
had been reported. Equation
(2.18) in Ref. \cite{Holas1} deserves fully appreciation,
for it is the first analytical expression obtained
for $Im \Pi_1 ({\bf k}, \omega)$ in terms of one-fold integral. 
Our expression, given as Eq. (\ref{Pi-final}) in Sec. IV, agrees 
numerically with Eq. (2.18) of Ref. \cite{Holas1}. The correctness of both
of them thus should be beyond doubt. It can be hardly denied that the method 
invented to obtain Eq. (2.18) in Ref. \cite{Holas1} is ingenious.
Our expression also in terms of one-fold integral has in contrast the 
character of simplicity. 
It is also the belief of the present author that this expression has been
obtained in optimal way and the derivation is more or less straightforward.
Overall, the exchange contribution included in the dielectric function 
makes a significant improvement over the random-phase approximation
(RPA), as had already been shown in Ref. \cite{Holas1} in several important 
respects. This will get full confirmation in this series of papers. We must 
further mention that the singular behavior of $\Pi_1 ({\bf k}, \omega)$ near
the characteristic frequencies $\omega_s = (\hbar /2m)|\pm k_Fk +k^2/2| $,
which had been elucidated in Ref. \cite{Holas1} and apparently
had made some negative impression of the many-body perturbation
theory on those authors \cite{Holas2}, is also
confirmed. Indeed explicit expressions of both of the discontinuity jump
of $Im \Pi_1 ({\bf k}, \omega)$ at $\omega= \omega_s$ and the 
corresponding logarithmic divergence there of its real counterpart are 
obtained in this paper, which are  presented in Sec. V.

In a series of papers, Brosens et al \cite{Brosens} investigated 
the local field correction to the RPA. 
They calculated 
the property 
\begin{eqnarray}
G({\bf k}, \omega)=
-v^{-1}(k) \Pi_1 ({\bf k}, \omega)/\Pi_0^2 ({\bf k}, \omega)
\end{eqnarray}
as an approximation to the local field factor \cite{Hubbard}. This property
is surely not the local field factor
including the exact exchange contribution, a fact evidently appreciated
by those authors.
The latter is instead
[according to Eq. (2) in I]  
\begin{equation}
G({\bf k}, \omega) =v(k)^{-1} \biggl [
\frac{1}{\Pi_0 ({\bf k}, \omega)+\Pi_1 ({\bf k}, \omega)}
-\frac{1}{\Pi_0 ({\bf k}, \omega)} \biggr ].
\end{equation}
They apparently had
never elucidated however, for the benifit of readers, that 
their approximation, obtained by them from
the dynamic-exchange decoupling in the equation of motion for 
the Wigner distribution function, could be also obtained as
an (sub-exchange in the sense explained above) approximation in 
the perturbation theory. (See also
the comments made in Ref. \cite{Holas1} on the earlier ones
of the series papers by Brosens et al.) They did
point out definitely that several forms obtained before and after them
\cite{Rajagopal,Tripathy} were very close to
or virtually identical to theirs. The relation between the theory
of Rajagopal \cite{Rajagopal} and that by Tripathy and Mandal \cite{Tripathy} 
was also pointed out
in Ref. \cite{Tripathy}. Tripathy and Mandal further elucidated the
relation between their theory and that proposed in Ref. \cite{Toigo}. A critical
analysis of the relation of the latter (in the static case) to
the first order theory was given earlier in Ref. \cite{Rasolt}.
Finally we wish to mention that Richardson 
and Ashcroft \cite{Richardson} also had obtained an analytical expression
for $\Pi_1 ({\bf k}, \omega)$ but with $\omega$ to be imaginary.
Investigations beyond the first order had also been attempted in general,
in Refs. \cite{Holas2,Gasser,Richardson} for instance, but mainly in limiting cases, 
in Refs. \cite{DuBois1,Glick2,Ninham,Hasegawa,Holas1,Holas3,Nifosi} 
again for instance.

Expression (\ref{Pi-final})
together with (\ref{H}) for $Im \Pi_1 (k, \omega)$ is the main result
of this paper. [We remind the reader that $Im \Pi ({\bf k}, \omega)$ 
determines fully $\Pi ({\bf k}, \omega)$, for 
its real conjugation can be
determined from it via the dispersion relation.]
The aim of this series of papers is to achieve
a (relatively speaking) complete and final
understanding of the role of the exchange contribution in the
dielectric function, taking advantage of the explicit form
of expression (\ref{Pi-final}) and that for $\Pi_1 ({\bf k}, 0)$
(Eq. (3) in I \cite{Qian,Engel}).
As an example, we mention that
it has been traditionally believed that $Im \Pi_1 ({\bf k}, \omega)$ 
has the limiting form of $\sim \omega$ for 
small $\omega$ \cite{Hasegawa,Nifosi,Vignale}.
In fact, it was claimed by Mahan \cite{Mahan} and has been commonly 
accepted that this must be true also for $Im \epsilon ({\bf k}, \omega)$,
the imaginary part of $\epsilon ({\bf k}, \omega)$,
in general. We find that this is not the case and 
$Im \Pi_1 ({\bf k}, \omega)$ actually has the limiting form
of $\sim \omega \ln \omega$, (details of which will be presented
in a subsequent paper.) 
The deep subtlety of many-body effects often reveals itself against 
our intuitive understanding, and does so most definitely and
convincingly in the perturbation theory indeed.
We end the introduction by further remarking that calculations
in the many-body perturbation theory are conventionally
known to be notoriously complicated. In this sense, our
expression appears quite simple.
The derivation to obtain it
has also been carried out in a quite manageable manner. Perhaps this is an
enlightening revelation about the many-body perturbation theory.

We give our derivation in Sec. III, after presenting the starting formalism 
in Sec. II.
 
\section{Starting formalism}

The Feynman-diagrammatically obtained expression for $\Pi_1 ({\bf k}, \omega)$
has been shown as Eq. (4) in I. It is, as is well known, the sum of 
two contributions:
\begin{equation}     \label{SE+Ex}
\Pi_1 ({\bf k}, \omega) =\Pi_1^{SE} ({\bf k}, \omega)
+\Pi_1^{Ex} ({\bf k}, \omega);
\end{equation}
$\Pi_1^{SE} ({\bf k}, \omega)$ and $\Pi_1^{Ex} ({\bf k}, \omega)$ arise,
respectively, from the self-energy diagrams and the exchange diagram.
We put down below the explicit expressions for them:
\begin{equation}
\Pi_1^{SE} ({\bf k}, \omega) =\frac{2}{\hbar^2}
\int  \frac{d{\bf p}}{(2 \pi)^3} \frac{d{\bf p}'}{(2 \pi)^3}
v({\bf p}- {\bf p}')
\frac{(n_{\bf p} -n_{{\bf p}+{\bf k}})(n_{{\bf p}'}
-n_{{\bf p}'+{\bf k}})}
{[\omega +\omega_{\bf p} -\omega_{{\bf p}+{\bf k}} +i0^+]^2} ,
\end{equation}
and
\begin{equation}
\Pi_1^{Ex} ({\bf k}, \omega) =-\frac{2}{\hbar^2}
\int  \frac{d{\bf p}}{(2 \pi)^3} \frac{d{\bf p}'}{(2 \pi)^3}
v({\bf p}- {\bf p}')
\frac{(n_{\bf p} -n_{{\bf p}+{\bf k}})(n_{{\bf p}'}
-n_{{\bf p}'+{\bf k}})}
{[\omega +\omega_{\bf p} -\omega_{{\bf p}+{\bf k}} +i0^+]
[\omega +\omega_{{\bf p}'}-\omega_{{\bf p}'+{\bf k}} +i0^+]}.
\end{equation}
(See also Refs. \cite{DuBois,Holas1,Geldart}.) The notations in this paper all follow I,
and here we have explicitly written $\hbar$.
With some manipulation, $\Pi_1^{SE} ({\bf k}, \omega)$ can be cast
in the following form:
\begin{eqnarray}  \label{sect2a}
\Pi_1^{SE} ({\bf k}, \omega) =  \frac{2m^2}{(2 \pi)^6 \hbar^2} \int d {\bf p}
\int d {\bf p}'
n_{{\bf p}-{\bf k}/2} n_{{\bf p}'-{\bf k}/2}
[v({\bf p}- {\bf p}')-v({\bf p}+ {\bf p}')]  \nonumber \\
\biggl [ \frac{1}{(m \omega - \hbar {\bf p} \cdot {\bf k}+i0^+)^2}
+\frac{1}{(m \omega + \hbar {\bf p} \cdot {\bf k}+i0^+)^2} \biggr ],
\end{eqnarray}
and $\Pi_1^{Ex} ({\bf k}, \omega)$:
\begin{eqnarray}  \label{sect2b}
 \Pi_1^{Ex} ({\bf k}, \omega) =  
&& - \frac{2m^2}{(2 \pi)^6 \hbar^2} \int d {\bf p}
\int d {\bf p}'
n_{{\bf p}-{\bf k}/2} n_{{\bf p}'-{\bf k}/2}  \nonumber \\
&& \biggl [ v({\bf p} - {\bf p}')
 \biggl (  \frac{1}{(m \omega - \hbar {\bf p} \cdot {\bf k}+i0^+)
(m \omega - \hbar {\bf p}' \cdot {\bf k}+i0^+)}     \nonumber \\
&&~~~~~~~~~~~~~~+\frac{1}{(m \omega + \hbar {\bf p} \cdot {\bf k}+i0^+)
(m \omega + \hbar {\bf p}' \cdot {\bf k}+i0^+)} \biggr )  \nonumber \\
&& -v({\bf p} + {\bf p}')
 \biggl ( \frac{1}{(m \omega - \hbar {\bf p} \cdot {\bf k}+i0^+)
(m \omega + \hbar {\bf p}' \cdot {\bf k}+i0^+)}           \nonumber \\
&&~~~~~~~~~~~~~~+\frac{1}{(m \omega + \hbar {\bf p} \cdot {\bf k}+i0^+)
(m \omega - \hbar {\bf p}' \cdot {\bf k}+i0^+)} \biggr ) \biggr ] .   
\end{eqnarray}
The imaginary parts of them can be obtained , respectively,
as
\begin{eqnarray} \label{Def-SE} 
Im \Pi_1^{SE} ({\bf k}, \omega) =&& 
\frac{m}{(2 \pi)^5 \hbar^2 } \frac{\partial}{\partial \omega}
 \int d {\bf p} \int d {\bf p}'
n_{{\bf p}-{\bf k}/2} n_{{\bf p}'-{\bf k}/2}   \nonumber \\
&& [v({\bf p}- {\bf p}')-v({\bf p}+ {\bf p}')]
[\delta (m \omega - \hbar {\bf p} \cdot {\bf k})
+\delta (m \omega + \hbar {\bf p} \cdot {\bf k})]  , 
\end{eqnarray}
and
\begin{eqnarray} \label{Def-Ex}  
Im \Pi_1^{Ex} ({\bf k}, \omega)  &=& \frac{2 m^2}{(2 \pi)^5 \hbar^2}  
 \int d {\bf p} \int d {\bf p}'
n_{{\bf p}-{\bf k}/2} n_{{\bf p}'-{\bf k}/2}   \nonumber \\
&~& \biggl [ v({\bf p}- {\bf p}')
\biggl ( \frac{1}{m \omega - \hbar {\bf p}' \cdot {\bf k}}
\delta (m \omega - \hbar {\bf p} \cdot {\bf k})
+\frac{1}{m \omega + \hbar {\bf p}' \cdot {\bf k}}
\delta (m \omega + \hbar {\bf p} \cdot {\bf k}) \biggr )  \nonumber \\
&-& v({\bf p}+ {\bf p}')
\biggl ( \frac{1}{m \omega - \hbar {\bf p}' \cdot {\bf k}}
\delta (m \omega + \hbar {\bf p} \cdot {\bf k})
+\frac{1}{m \omega + \hbar {\bf p}' \cdot {\bf k}}
\delta (m \omega - \hbar {\bf p} \cdot {\bf k}) \biggr ) \biggr ] . \nonumber \\
~~
\end{eqnarray}
These forms serve our purpose best.

\section{ Derivation}

\subsection{ $Im \Pi_1^{SE} ({\bf k}, \omega)$ }

The property
$\Pi_1 ({\bf k}, \omega)$ depends only on the magnitude of ${\bf k}$ 
in a uniform system,
so it may be written as $\Pi_1 ( k, \omega)$. We first define 
a dimensionless quantity: $\Omega=m \omega/\hbar k_F^2$. 
From now on throughout the paper we put $k$ in units of $k_F$, i.e.,
$k$ will always be dimensionless.

The computation for $Im \Pi_1^{SE} (k, \omega)$ can be made 
very simple. The integral over the variable ${\bf p}'$ in Eq. (\ref{Def-SE})
can be carried out first, which leads to
\begin{eqnarray}  \label{SE-1}
Im \Pi_1^{SE} (k, \omega) = \frac{m^2e^2}{2 \pi^2 \hbar^4} 
\frac{\partial}{\partial \Omega}
&~& \int_{-a}^b dz \int_0^\lambda dx
 [\delta(\Omega -kz) +\delta(\Omega +kz) ]     \nonumber  \\   
&~& [F(\sqrt{z^2 +x -kz +k^2/4}) 
-F(\sqrt{z^2 +x +kz +k^2/4}) ] ,
\end{eqnarray}
where
\begin{eqnarray}
F (q) = \frac{1}{4 \pi} \int d {\bf p} 
\frac{n_{\bf p}}{|{\bf p} -{\bf q}|^2} . 
\end{eqnarray}
Explicitly,
\begin{eqnarray}
F (q) = \frac{1}{2} + \frac{1 -q^2}{4q}  
\ln  \biggl |\frac{1+q}{1-q} \biggr |. 
\end{eqnarray}
We mention once again that the notations here follow I.
The integration over $z$ in Eq. (\ref{SE-1}) is trivial. After performing it,
one gets
\begin{eqnarray} \label{SE-2}
Im \Pi_1^{SE} ( k, \omega) = \frac{m^2 e^2}{2 \pi^2 \hbar^4}\frac{1}{k^2}
&\bigg [& \theta \{(b-\Omega /k)(a+\Omega/k) \} 
H^{SE} (k, \Omega/k)     \nonumber \\
 &~& - \theta  \{(b+\Omega /k)(a-\Omega/k) \} 
H^{SE} (k, -\Omega/k) \biggr ] ,
\end{eqnarray}
with
\begin{eqnarray}
H^{SE} (k, z) =  \frac{\partial}{\partial z} 
\int_0^\lambda dx  [F(\sqrt{x -\lambda +1 })-F(\sqrt{x -\lambda +1 + 2kz}) ] .
\end{eqnarray}
The $H^{SE} (k, z)$ in the preceding equation can be readily refined into
\begin{eqnarray}
H^{SE} (k, z) = (k-2z)F(\sqrt{-\lambda +1}) +(k+2z) F(\sqrt{-\lambda +1+2kz})
-2k F(\sqrt{1+2kz}) . 
\end{eqnarray}
Explicitly,
\begin{eqnarray}     \label{SE-final}
H^{SE} (k, z) = \frac{1}{2}  
\biggl [ 2 k^2 z \frac{1}{\sqrt{C_0}} Y(z)
-  \lambda W_1 (z)  -  {\tilde \lambda} W_2 (z) \biggr ].
\end{eqnarray}
In Eq. (\ref{SE-final}) we have introduced (newly) the
symbol ${\tilde \lambda}=(b+z)(a-z)$.

\subsection{ $Im \Pi_1^{Ex} ({\bf k}, \omega)$ }

Our labor lies mainly in the evaluation of $Im \Pi_1^{Ex} (k, \omega)$
expressed in (\ref{Def-Ex}).
Following paper I, we first carry out the integrals over the azimuthal 
angular variables of ${\bf p}$ and ${\bf p}'$. After that, we obtain
\begin{eqnarray}
Im \Pi_1^{Ex} ( k, \omega)
=&& \frac{m^2 e^2}{4 \pi^2 \hbar^4}  
\int_{-a} \int^b dz dz' \biggl [ 
\biggl \{\frac{1}{\Omega -k z'} \delta (\Omega - k z)
+\frac{1}{\Omega +k z'} \delta (\Omega + k z) \biggr \} 
L(\beta^2)                                         \nonumber \\
&&~~~~~~~~~~~~~~~~~~
-\biggl \{ \frac{1}{\Omega -k z'} \delta (\Omega + k z)
+\frac{1}{\Omega +k z'} \delta (\Omega - k z) \biggr \} 
L(\alpha^2) \biggr ] .
\end{eqnarray}
We then, taking advantage of the presence of the 
$\delta- $ function, reduce the two-fold integral
to one-fold. The $Im \Pi_1^{Ex} ( k, \omega)$ becomes thus
\begin{eqnarray}     \label{Ex}
Im \Pi_1^{Ex} ( k, \omega)= - \frac{m^2e^2}{4 \pi^2 \hbar^4} \frac{1}{k^2}
&\bigg [&\theta \{(b-\Omega /k)(a+\Omega/k) \} 
H^{Ex} (k, \Omega/k)    \nonumber \\
&~& - \theta \{(b+\Omega /k)(a-\Omega/k) \} 
H^{Ex} (k, -\Omega/k) ] ,
\end{eqnarray}
with the function $H^{Ex} (k, z)$ defined as
\begin{eqnarray} \label{Def-HEx}
H^{Ex} (k, z)= \int_{-a}^b dz'
\biggl [ \frac{1}{\alpha} L(\alpha^2) -\frac{1}{\beta} L(\beta^2) \biggr ] .
\end{eqnarray}
The function $L$ has been given in Eq. (9) in I and in Ref. \cite{Glasser}. 
There are several components in it, and we separate them in
the evaluation of the integral in Eq. (\ref{Def-HEx}). Accordingly we write 
$H^{Ex} (k, z)$ in the following manner:
\begin{eqnarray} \label{HEx-2}
H^{Ex} (k, z)= H_0^{Ex} (k, z) + H_1^{Ex} (k, z)+H_{23}^{Ex} (k, z) ,
\end{eqnarray}
with
\begin{eqnarray}
H_0^{Ex} (k, z)= \int_{-a}^b dz' z' \bigg [ (\lambda + \lambda')
(2 \ln 2 +1) \frac{1}{\alpha \beta} - 1 \biggr ] ,
\end{eqnarray}
\begin{eqnarray}
H_1^{Ex} (k, z)= \int_{-a}^b dz' \bigg [ \frac{1}{\alpha}
\sqrt{R(z, z')} - \frac{1}{\beta} |\beta |\biggr ] ,
\end{eqnarray}
and
\begin{eqnarray} \label{Def-H23}
H_{23}^{Ex} (k, z) &=& \int_{-a}^b dz' 
 \biggl [  \biggl ( \frac{1}{\alpha}-\frac{1}{\beta} \biggr ) 
 \lambda \ln |4 \lambda|
- 2\lambda' \biggl( \frac{1}{\alpha} \ln |\alpha | 
- \frac{1}{\beta} \ln |\beta| \biggr )                 \nonumber \\
&-& \frac{1}{\alpha} \biggl ( \lambda \ln | \alpha^2 + \lambda' 
-\lambda -2 \sqrt{R(z, z')} |
- \lambda' \ln | \alpha^2 - \lambda' 
+\lambda +2 \sqrt{R(z, z')} | \biggr )   \nonumber \\
&+& \frac{1}{\beta} \biggl ( \lambda \ln | \beta (k-2z) + 2|\beta| |
- \lambda' \ln | \beta(k-2z') + 2|\beta| | \biggr ) \biggr ] .     
\end{eqnarray}
The two terms of $J_2$ and $J_3$ \cite{Glasser} in Eq. (11) of I were combined, 
for the simplicity of the computation, into one term [denoted as $J_{23}$ in 
Eq. (43) there]. The $H_{23}^{Ex} (k, z)$ here follows suit.

The evaluation for $H_0^{Ex} (k, z)$ and $H_1^{Ex} (k, z)$ is a routine
job. It is quite straightforward to get the following result:
\begin{eqnarray}      \label{H0}
H_0^{Ex} (k, z)=  
(2 \ln 2 +1) [ \lambda W_1 (z) + (ab-z^2) W_2(z) - k ] - k ,
\end{eqnarray}
and
\begin{eqnarray}     \label{H1}
H_1^{Ex} (k, z)=                                                
k [2 - z W_2(z) - z (2z +k) C_0^{-1/2} Y(z) ] .
\end{eqnarray}

We next attack $H_{23}^{Ex} (k, z)$. With a little
algebra,  we rewrite Eq. (\ref{Def-H23}) in
the following form:
\begin{eqnarray}    \label{H23}
H_{23}^{Ex} (k, z)= - [ W_1(z) + W_2(z) ] \lambda \ln |4 \lambda |
-2 \zeta_1 (z) - \zeta_1 (-z) + \zeta_2 (z) -\zeta_3(z) , 
\end{eqnarray}   
with
\begin{eqnarray}
\zeta_1 (z) = \int_{-a}^b dz' \frac{\lambda'}{\alpha} 
\ln | \alpha | ,
\end{eqnarray}
\begin{eqnarray}
\zeta_2 (z) = \int_{-a}^b dz' \frac{1}{\beta}
\bigg [ \lambda \ln | \beta(k-2z) + 2|\beta| |
- \lambda' \ln | k-2z' + 2 \beta /|\beta| |  \biggr ] ,
\end{eqnarray}
and
\begin{eqnarray}   \label{zeta-3-Def}
\zeta_3 (z) = \int_{-a}^b dz' \frac{1}{\alpha}
\bigg [ \lambda \ln | \alpha^2 + \lambda' -\lambda -2 \sqrt{R(z, z')} |
- \lambda' \ln | \alpha^2 - \lambda' +\lambda + 2 \sqrt{R(z, z')} | \biggr ] .
\end{eqnarray}
The reader should not confuse the functions $\zeta_n (z)$ here with the Riemann's
function $\zeta (n)$ that appeared in I.
The evaluation for $\zeta_1(z)$ and $\zeta_2(z)$ is 
a little tedious but clearly straightforward. We thus present only the results:
\begin{eqnarray}    \label{zeta-1}
\zeta_1 (z) = 
\frac{1}{2}  [ (2z+k)(\ln |{\tilde \lambda}| -3 )
+ \{ {\tilde \lambda} (3 -\ln |{\tilde \lambda}|) -2 \} W_2 (z)  ] ,
\end{eqnarray}
and
\begin{eqnarray}    \label{zeta-2}
\zeta_2 (z) = 
\frac{1}{2} [\lambda (\ln \lambda -2 \ln2 -3) +2 ] W_1 (z) 
+\frac{1}{2}(2z -k) (\ln \lambda -6 \ln2 +3) - \lambda v_1 (z) ,
\end{eqnarray}
where
\begin{eqnarray}
v_1 (z) = \int_{-a}^z dz' \frac{1}{z' -b } \ln \beta     
+\int_{z}^b dz' \frac{1}{z' +a } \ln \beta .
\end{eqnarray}

We now turn to $\zeta_3 (z)$. We first rewrite Eq. (\ref{zeta-3-Def}) as
\begin{eqnarray}  \label{zeta-3}
\zeta_3 (z) = \lambda \zeta_{3a} (z) - \zeta_{3b} (z) ,
\end{eqnarray}
with 
\begin{eqnarray} \label{zeta-3a}
\zeta_{3a} (z) =  \int_{-a}^b dz' \frac{1}{\alpha}
\ln | \alpha^2 + \lambda' -\lambda -2 \sqrt{R(z, z')} | ,
\end{eqnarray}
and
\begin{eqnarray}  \label{zeta-3b}
\zeta_{3b} (z) = \int_{-a}^b dz' \frac{\lambda '}{\alpha}
\ln | \alpha^2 - \lambda' +\lambda + 2 \sqrt{R(z, z')} | .
\end{eqnarray}
The evaluation of $\zeta_{3a} (z)$ is also a routine job. It can be
effected with partial integration. One gets in this manner
\begin{eqnarray} \label{3a}
\zeta_{3a} (z) = 
- W_2 (z) \ln |2 \lambda | - \int_{-a}^b dz' D_1 (z, z') \ln |\alpha | ,
\end{eqnarray} 
where
\begin{eqnarray}
D_1 (z, z') = \frac{\partial }{\partial z'}
\ln | \alpha^2 + \lambda' - \lambda - 2 \sqrt{R(z, z')} | .
\end{eqnarray}
Explicitly,
\begin{eqnarray}
D_1 (z, z') =
\frac{1}{\alpha} \biggl [1 - \frac{kz}{\sqrt{R(z, z')}} \biggr ] .
\end{eqnarray}
The equation (\ref{3a}) can now be readily refined into
\begin{eqnarray} \label{zeta-3a-final}
\zeta_{3a} (z) = 
\frac{1}{2} [ \ln |{\tilde \lambda}|
- 2 \ln |2 \lambda | ] W_2 (z) + kz v_2 (z) ,
\end{eqnarray}
with
\begin{eqnarray} \label{v2}
v_2 (z) = \int_{-a}^b dz' \frac{1}{\alpha \sqrt{R(z, z')} }
\ln |\alpha | .
\end{eqnarray}
The integral on the right hand side of Eq. (\ref{zeta-3b}) can also be 
effected with partial integration. To this end, we employ the following
identity:
\begin{eqnarray} 
\frac{\lambda '}{\alpha} dz' = \frac{1}{2} 
d [2 {\tilde \lambda} \ln |\alpha| + \lambda ' + (2z+k) \alpha ],  
\end{eqnarray} 
with the symbol $d$ here denoting the differential operating only on the
variable $z'$. We perform in this manner the partial integration and get
for $\zeta_{3b} (z)$:
\begin{eqnarray}         \label{zeta-3b-1}
\zeta_{3b} (z) &=& \frac{1}{2} (2z +k) 
[ 4 \ln 2 +(b+z) \ln r_1 +(a-z) \ln r_2 ]
+ {\tilde \lambda} [ -2 W_2(z) \ln 2 + v_0 (z)]         \nonumber \\
&~& - \frac{1}{2} \int_{-a}^bd z' 
 [2 {\tilde \lambda} \ln |\alpha| + \lambda' + (2z+k) \alpha ] D_2 (z, z') ,
\end{eqnarray}
where
\begin{eqnarray}   \label{v0}
v_0 (z) = \ln |b+z| \ln r_1 - \ln |a-z| \ln r_2 ,
\end{eqnarray}
and
\begin{eqnarray}
D_2(z, z') = \frac{\partial }{\partial z'}
\ln | \alpha^2 - \lambda' +\lambda + 2 \sqrt{R(z, z')} | .
\end{eqnarray}
We have introduced the following symbols:
\begin{eqnarray}    \label{r1-r2}
r_1 = \sqrt{R(z, b)} ,  ~~~~~~ r_2 =\sqrt{R(z, -a)} 
\end{eqnarray}
in Eqs. (\ref{zeta-3b-1}) and (\ref{v0}).
Explicitly
\begin{eqnarray}   \label{D2}
D_2(z, z') =  
\frac{1}{\alpha} \biggl [1 - \frac{kz}{\sqrt{R(z, z')}} \biggr ]
- \frac{1}{2(b-z')} \biggl [ 1- \frac{r_1}{\sqrt{R(z, z')}} \biggr ]  
+ \frac{1}{2(a+z')} \biggl [1 - \frac{r_2}{\sqrt{R(z, z')}} \biggr ] . 
                                                              \nonumber \\ 
\end{eqnarray}
Making the use of Eq. (\ref{D2}) in Eq. (\ref{zeta-3b-1}), the expression
for $\zeta_{3b} (z)$ 
can be organized in the form:
\begin{eqnarray}         \label{zeta-3b-final}
\zeta_{3b} (z)=&&  [ (2z +k)
\{ 4 \ln 2 +(b+z) \ln r_1 +(a-z) \ln r_2 \}
+ {\tilde \lambda} \{ -4 W_2(z) \ln 2 + 2v_0 (z)     \nonumber \\
&& +W_2(z) \ln |{\tilde \lambda}| + 2kz v_2 (z) + v_3 (z) \}
- {\bar \zeta}_{3b} (z)  ]/2 ,
\end{eqnarray}
where $v_3(z)$ and ${\bar \zeta}_{3b} (z)$
are defined as 
\begin{eqnarray}
v_3 (z) =  
\int_{-a}^b d z' \ln |\alpha|
\biggl [\frac{1}{b-z'} \biggl ( 1- \frac{r_1}{\sqrt{R(z, z')}} \biggr )     
+\frac{1}{a+z'} \biggl ( -1+ \frac{r_2}{\sqrt{R(z, z')}} \biggr ) \biggr ] ,
\end{eqnarray}
and
\begin{eqnarray}   \label{bar0}
{\bar \zeta}_{3b} (z)
=\int_{-a}^b dz'[\lambda' +(2z +k) \alpha ] D_2 (z, z') ,
\end{eqnarray}
respectively.
The integral on the right hand side of Eq. (\ref{bar0})
[with $D_2 (z, z')$ given explicitly in Eq. (\ref{D2})] is 
basic, although it looks somewhat tedious. With some algebra, we can put it
in the following form:
\begin{eqnarray}    \label{bar}
{\bar \zeta}_{3b} (z)=&&
-{\tilde \lambda}W_2 (z) -  (z/2) [ 3 (k^2 -1 ) +C_0 ] V_0 (z)
+C_0 V_1 (z) - kz {\tilde  \lambda} V_{-1} (z, -z)    \nonumber \\
&& -\frac{1}{2} (2z +k) \biggl [ - 10 + 4C_0 V_0 (z)
+ (z +b) \biggl ( \int_{-a}^b dz' \frac{1}{b-z'} 
+ r_1 V_{-1} (z, b) \biggr )     \nonumber \\
&& - (z -a) \biggl ( \int_{-a}^b dz' \frac{1}{a+z'} 
- r_2 V_{-1} (z, -a) \biggr ) \biggr ] ,
\end{eqnarray}
where
\begin{eqnarray}
V_n (z)=\int_{-a}^b dz' \frac{z'^n}{\sqrt{R(z, z')}} 
\end{eqnarray}
for $ n=0, 1$, and 
\begin{eqnarray}
V_{-1} (z, x)=
\int_{-a}^b dz' \frac{1}{z'-x} \frac{1}{\sqrt{R(z, z')}} .
\end{eqnarray}
Explicitly \cite{Qian},
\begin{eqnarray}
V_0 (z)=2 C_0^{-1/2} Y(z) ,
~~~ ~~~ ~~~~
V_1 (z)= [2z+k - z(kz + 1 -k^2/2)  V_0 (k, z)] C_0^{-1} ,
\end{eqnarray}
and
\begin{eqnarray}
V_{-1} (z, -z)= - W_2 (z) /kz .
\end{eqnarray}
We note that the sum in each of the big curve bracket in Eq. (\ref{bar})
is well defined, though their respective components are not. The reader excuses us 
for the sake of a compact presentation. Indeed,
\begin{eqnarray}
\int_{-a}^b dz' \frac{1}{b-z'}
+ r_1 V_{-1} (z, b) =  2 \ln  \biggl | \frac{kz}{r_1} \biggr | ,
\end{eqnarray}
and 
\begin{eqnarray}
\int_{-a}^b dz' \frac{1}{a+ z'}
- r_2 V_{-1} (z, -a) =  2 \ln  \biggl | \frac{kz}{r_2} \biggr | .  
\end{eqnarray}
In virtue of the foregoing results, the integral on the right hand side of 
Eq. (\ref{bar}) has now been fully 
carried out. The final result for ${\bar \zeta}_{3b} (z)$ can 
after refinement be written as
\begin{eqnarray}   \label{bar-final}
{\bar \zeta}_{3b} (z) =
(2z+k) [ 6 - 2(3kz+1)C_0^{-1/2} Y(z) +(z+b) \ln r_1 
- (z-a) \ln r_2 - 2 \ln |kz| ] .
\end{eqnarray}
The substitution of Eq. (\ref{bar-final}) into Eq. (\ref{zeta-3b-final}) 
will give the result for $\zeta_{3b} (z)$. Further substitution of thus
obtained result for $\zeta_{3b} (z)$ and the previously obtained one
for $\zeta_{3a} (z)$ in Eq. (\ref{zeta-3a-final}) into 
Eq. (\ref{zeta-3}) then yields the final result for $\zeta_3 (z)$, which 
turns out to be
\begin{eqnarray}     \label{zeta-3-final}
\zeta_3 (z) =&~& [kz \ln |{\tilde \lambda}| - \lambda \ln | 2 \lambda |
+ 2 {\tilde \lambda} \ln 2 ] W_2 (z) +2 k^2 z^2 v_2 (z)  
- ({\tilde \lambda}/2) [2 v_0 (z) + v_3 (z)]      \nonumber \\
&-& (2z +k) [ 2 \ln 2 -3 + (3kz +1) C_0^{-1/2} Y(z) + \ln |kz| ] .
\end{eqnarray}

One then substitutes $\zeta_1 (z)$ expressed in Eq. (\ref{zeta-1}), 
$\zeta_2 (z)$ in Eq. (\ref{zeta-2}), and $\zeta_3 (z)$ in the above
equation into Eq. (\ref{H23}) to get
\begin{eqnarray}   \label{H23-final}
H_{23}^{Ex} (k , z)=&& (-2z +5k) \ln 2 + \mu_1 (k, z) + \mu_1 (k, -z) 
- 2 (1+ \ln 2) [\lambda W_1 (z) + {\tilde \lambda} W_2 (z)]   \nonumber \\
&&- [2kz \ln 2 +(kz -{\tilde \lambda} ) \ln |{\tilde \lambda}|] W_2 (z)     
- \lambda v_1 (z) 
+  ( {\tilde \lambda} /2) [2 v_0 (z) +v_3 (z) ]  \nonumber \\
&& +(2z +k) [(3kz +1) C_0^{-1/2} Y(z) + \ln |kz|] - 2k^2z^2 v_2 (z) ,
\end{eqnarray}
where
\begin{eqnarray}
\mu_1 (k, z) = (2z -k) \ln \lambda + [2- (1+\ln 2 ) \lambda ] W_1 (z) .
\end{eqnarray}
One then advances further to add [according to Eq. (\ref{HEx-2})] 
$H_0^{Ex} (k , z)$ of 
Eq. (\ref{H0}), 
$H_1^{Ex} (k , z)$ of Eq. (\ref{H1}), and 
$H_{23}^{Ex} (k , z)$ of Eq. (\ref{H23-final}) to
get the result for $H^{Ex} (k , z)$ which can be in the final form written as
\begin{eqnarray}   \label{Ex-final}
H^{Ex} (k , z)=&& (-2z +3k) \ln 2 + \mu_1 (k, z) +  \mu_1 (k, -z)
+ ({\tilde \lambda} -kz) W_2 (z) \ln |{\tilde \lambda}|    \nonumber \\
&& - \lambda W_1 (z) - {\tilde \lambda} W_2 (z)    
- \lambda v_1 (z) + ({\tilde \lambda /2)} [2 v_0 (z) +v_3 (z) ]
\nonumber \\
&& + (2z +k) [\sqrt{C_0} Y(z) + \ln |kz|] - 2k^2z^2 v_2 (z) .
\end{eqnarray}

\section{Analytical result}

In virtue of Eq. (\ref{SE+Ex}), $Im \Pi_1 (k, \omega)$ can be obtained from
Eq. (\ref{SE-2}) and Eq. (\ref{Ex}) as
\begin{eqnarray}  \label{Pi-final}
Im \Pi_1 (k, \omega) = \frac{m^2e^2}{(2 \pi)^2 \hbar^4} \frac{1}{k^2}
[ \theta  (1-\nu_+^2) H (k, \Omega/k)     
 - \theta  (1-\nu_-^2) H (k, -\Omega/k) ] ,  
\end{eqnarray}
with
$\nu_+= \Omega/k -  k/2$ , 
$\nu_-= -\Omega/k -  k/2$ ,
and
\begin{eqnarray}    
H (k, z)= 2 H^{SE} (k, z) -  H^{Ex} (k, z) .
\end{eqnarray}
The substitution from Eqs. (\ref{SE-final}) and (\ref{Ex-final}) will give
the result for $H (k, z)$, which we then further refine into the following form:
\begin{eqnarray}    \label{H}
H (k, z)= && - (2zC_0 +k) C_0^{-1/2} Y(z) + (2z - 3k) \ln 2
- (k +2z) \ln |kz|                                           \nonumber \\              
&& + (kz - {\tilde \lambda} ) \ln |{\tilde \lambda}| W_2 (z)
- \mu_1 (k, z) - \mu_1 (k, -z) - \mu_2 (k, z) +\mu_2 (-k , -z)    \nonumber \\
&&+ 2 k^2 z^2 \int_{-a}^b d z' \frac{1}{\alpha \sqrt{R(z, z')}} \ln |\alpha| , 
\end{eqnarray}
where
\begin{eqnarray}    \label{mu2}
\mu_2 (k, z) = && {\tilde \lambda} \ln |z+b| \ln |(k+1)z +b|
-\lambda \int_z^b dz' \frac{1}{z' +a} \ln |\beta|           \nonumber \\
&& +\frac{1}{2} {\tilde \lambda} \int_{-a}^b dz' \frac{1}{b-z'} 
\biggl [ 1 - \frac{(k+1)z +b}{\sqrt{R(z, z')}} \biggr ] \ln |\alpha| .
\end{eqnarray}

\section{Singularity of $\Pi_1 (k, \omega)$ at $\omega = \omega_s$ }

One can immediately see that $Im \Pi_1 ( k, \omega)$ has the same
nonvanishing region as
$Im \Pi_0 ( k, \omega)$, the Lindhard function \cite{Lindhard,Fetter,Mahan}. 
In other words, the region of the single particle-hole continuum
remains unchanged with the inclusion of the exchange contribution.
The long-wavelength plasmon which has zero linewidth in RPA
up to wavevector $k_c$, at which the damping sets in, accordingly 
remains up to $k_c$ infinitely robust against exchange effect.
This truth has been recognized before \cite{DuBois1,Glick2,Ninham,
Hasegawa,Holas1,Holas3,Gasser,Nifosi}. Such a distinctly 
drawn conclusion, if 
understood in an appropriate manner, must also be 
appreciated as one of the merits
of the perturbation theory.

While $Im \Pi_0 ( k, \omega)$ approaches to zero on the edge of the
single particle-hole continuum,
$Im \Pi_1 (k, \omega)$ shows a discontinuity jump there.
In other
words, $\Pi_1 (k, \omega)$ exhibits singular behavior
at $\omega = \omega_{s}$ with $\omega_s = (\hbar k_F^2/2m)|\pm k +k^2/2| $.
This singularity
was noticed by Glick \cite{Glick} before Holas et al \cite{Holas1}, and
also by Awa et al \cite{Awa} after them, and Holas et al had made
the most elaborate investigation of it. In fact, all of the three groups of
authors had adopted a similar approach in order to remove it.
The jump discontinuity, defined as $\bigtriangleup _s  (k)
= Im \Pi_1 ( k, \omega_{s}+ 0^+) - Im \Pi_1 ( k, \omega_{s}- 0^+)$ can be 
explicitly calculated by the use of Eq. (\ref{Pi-final}). For
$\omega_{s} = (\hbar k_F^2/2m)(k+k^2/2) $,
\begin{eqnarray}
\bigtriangleup _s  (k) = 
\frac{m^2 e^2 }{2 \pi^2 \hbar^4}
\frac{b}{k(1+k)} \ln \biggl |\frac{2b}{k} \biggr | ; \nonumber \\
\end{eqnarray}
and, for $\omega_{s} = (\hbar k_F^2/2m) |-k+k^2/2| $,
\begin{eqnarray}
\bigtriangleup _s  (k) = -
\frac{m^2 e^2} {2 \pi^2 \hbar^4}
\frac{a}{k(1-k)} \ln \biggl |\frac{2a}{k} \biggr | .
\end{eqnarray}
The discontinuity in $Im \Pi_1 ( k, \omega)$ gives rise to a logarithmic
divergence in $Re \Pi_1 ( k, \omega)$, which has the following form
(to the accuracy of the leading logarithmic order):
\begin{eqnarray} 
Re \Pi_1 ( k, \omega) =  \frac{1}{\pi} \bigtriangleup _s  (k) 
\ln |2m (\omega - \omega_s)/\hbar k_F^2 | 
\end{eqnarray}
for $\omega \to \omega_s$.

\end{document}